\documentclass[12pt]{iopart}
\usepackage{bm}
\usepackage{graphicx}
\usepackage{color}

\begin{document}

\title[] {Unusual Exchange Bias in Sr$_2$FeIrO$_6$/La$_{0.67}$Sr$_{0.33}$MnO$_3$ Multilayer}

\author{K C Kharkwal, Rachna Chaurasia and A K Pramanik}

\address{School of Physical Sciences, Jawaharlal Nehru University, New Delhi - 110067, India}

\eads{\mailto{akpramanik@mail.jnu.ac.in}}

\begin{abstract}
Here, we study an interface induced magnetic properties in 3$d$-5$d$ based multilayer made of La$_{0.67}$Sr$_{0.33}$MnO$_3$ and double perovskite Sr$_2$FeIrO$_6$, respectively. Bulk La$_{0.67}$Sr$_{0.33}$MnO$_3$ is metallic and shows ferromagnetic (FM) ordering above room temperature. In contrast, bulk Sr$_2$FeIrO$_6$, is an antiferromagnet (AFM) with N$\acute{e}$el temperature around 45 K ($T_N$) and exhibits an insulating behavior. Two set of multilayers have been grown on SrTiO$_3$ (100) crystal with varying thickness of FM layer. multilayer with equal thickness of La$_{0.67}$Sr$_{0.33}$MnO$_3$ and Sr$_2$FeIrO$_6$ ($\sim$ 10 nm) shows exchange bias (EB) effect both in conventionally field cooled (FC) as well as in zero field cooled (ZFC) magnetic hysteresis measurements which is rather unusual. The ZFC EB effect is weakened both with increasing maximum field during initial magnetization process at low temperature and with increasing temperature. Interestingly, multilayer with reduced thickness of La$_{0.67}$Sr$_{0.33}$MnO$_3$ ($\sim$ 5 nm) does not exhibit ZFC EB phenomenon, however, the FC EB effect is strengthened showing much higher value. We believe that an AFM type exchange coupling at interface and its evolution during initial application of magnetic field cause this unusual EB in present multilayers.
\end{abstract}


\maketitle

\section{Introduction}
Artificially constructed interface between two dissimilar materials is of particular interest since new exotic states of matter emerge at interface which are not observed in its bulk counterpart. The interface properties in oxide materials have been widely studied during last several years showing many interesting physical properties which include unusual electronic transport and magnetism, superconductivity, ferroelectricity, exchange bias effect, etc.\cite {Bert, Li, Gib, Grut, San, Nak, Gan, Rey, Ban, Dong, Oht} Oxides, in general, has strong electronic correlation and complex interplay between charge, spin, orbital and lattice degrees of freedom. Therefore, confinement of electrons in nearly 2-dimensional (D) regime and reconstruction of charge, spin, orbital and lattice parameters at the interface are believed to cause such interesting properties. \cite{Hwang,Chak,Zubko,Mann,Ham} 

Among the oxide materials, while though 3$d$/4$d$ based heterostructures have been extensively studied,\cite{He,Pad} the multilayers constituted with 3$d$/5$d$ based materials are very less explored. The 3$d$/5$d$ systems carry a special interest because in addition to conventional interfacial effect, these multilayers provide an ideal system to study an interplay between electron correlation ($U$) and spin-orbit coupling (SOC) effect.\cite{Pes,Sam} The 3$d$ oxides usually show large $U$ which is reduced in 5$d$ materials due to its extended character of $d$ orbitals. The 5$d$ oxides, on the other hand, exhibit reasonable SOC with its heavy elements, therefore relevant energies share a comparable scale in these materials. Among 5$d$ oxides, iridates have special interest. The high crystal field effect (CFE) splits $d$ orbitals in these materials into $t_{2g}$ and $e_{g}$ states where a strong SOC further splits the low lying $t_{2g}$ states into $J_{eff}$ = 3/2 quartet and $J_{eff}$ = 1/2 doublet.\cite{kim, kim1} This gives a magnetic Ir$^{4+}$ (5$d^5$) and non-magnetic Ir$^{5+}$ (5$d^4$) with $J_{eff}$ = 1/2 and 0 ground state, respectively. Hence, the electro-magnetic properties are largely tunable in iridates. Ir based oxides have recently been focused and therefore a detail investigation on both bulk as well as thin films are required to understand their properties.\cite {Nic, Yi, Oka}

Here, we have studied an interface induced magnetic properties through exchange bias (EB) phenomenon in multilayers composed of 3$d$ based La$_{0.67}$Sr$_{0.33}$MnO$_3$ (LSMO) and 3$d$-5$d$ based double perovskite (DP) Sr$_2$FeIrO$_6$ (SFIO) materials. The LSMO shows ferromagnetic (FM) ordering down to low temperature with transition temperature ($T_c$) above room temperature.\cite {Ett} The 3$d$-5$d$ based DP systems are of recent interest because complex interaction between $U$ and SOC can be studied in same DP materials. Our recent study shows bulk SFIO is insulating down to low temperature and exhibits prominent antiferromagnetic (AFM) ordering at temperature $\sim$ 45 K ($T_N$), however, a weak AFM ordering has also been observed around 120 K. Magnetism in SFIO is only realized through Fe$^{3+}$ (3$d^5$) channel as Ir$^{5+}$ (5$d^4$) appears to be nonmagnetic within the picture of strong SOC.\cite {Khark} The interface in present multilayer represent a meeting point of FM and AFM magnetic state with three different transition metals (Mn, Fe and Ir) which is rather uncommon.\cite {Fan,Jia, Yi, Oka} This FM/AFM based multilayer and interface is of further interest considering present developments in the field of AFM based spintronics and its applications.\cite{balt}

The EB effect manifest through shifting of magnetic hysteresis loop ($M(H)$) along the magnetic field axis when the system is cooled in magnetic field from above $T_N$, and an interface between FM and AFM state works a precursor. It is mostly believed that the process of field cooling induces an unidirectional FM anisotropy at the interface which causes EB effect. Apart from its fundamental interest, EB phenomenon has widely been studied with interest for technological applications such as, in magnetic storage, magnetic sensors, spintronics, etc.\cite{Goke,Nogues,Nog} Conventionally, $M(H)$ loop shifts to negative field direction when cooled in positive magnetic field or vice versa which is called as negative EB effect. There are, however, few reports which shows EB effect even when the system is cooled in zero magnetic field, which is known as zero field cooled (ZFC) EB or spontaneous EB effect.\cite{Wang,Saha,Murthy} Artificially designed multilayers are naturally the best choices for EB systems as not only an individual layer component can be chosen with particular magnetic state but the interface play crucial role here with a modified magnetic character. However, the technological challenges to observe the spin structure at the interface often impose difficulties to completely understand this EB phenomenon in multilayers.\cite{Meik,Stamps,Kiwi,Berko,Ohl,Wu}   

In this present study, we have deposited epitaxial multilayers of [SFIO/LSMO]$_3$ with different layer thickness on SrTiO$_3$ (100) single crystal substrate and investigated the detailed magnetic properties. With lowering in temperature, magnetization continuously increases following pattern of LSMO, however, below around 45 K ($T_N$ of SFIO) magnetization data show a large bifurcation between zero field cooled and field cooled process. Multilayer with equal layer thickness of LSMO and SFIO exhibit EB effect in both ZFC and FC $M(H)$ hysteresis loop, however, with decreasing thickness of FM LSMO the ZFC EB effect vanishes. The calculated EB field $H_{EB}$ has been found to decrease with increase of applied field at 5 K in ZFC $M(H)$ data. Further, both the coercive field  and the exchange bias both decreases with increasing temperature and the $H_{EB}$ almost vanishes when temperature increases to $T_N$ of SFIO. 

\begin{figure}
\centering
		\includegraphics[width=8cm]{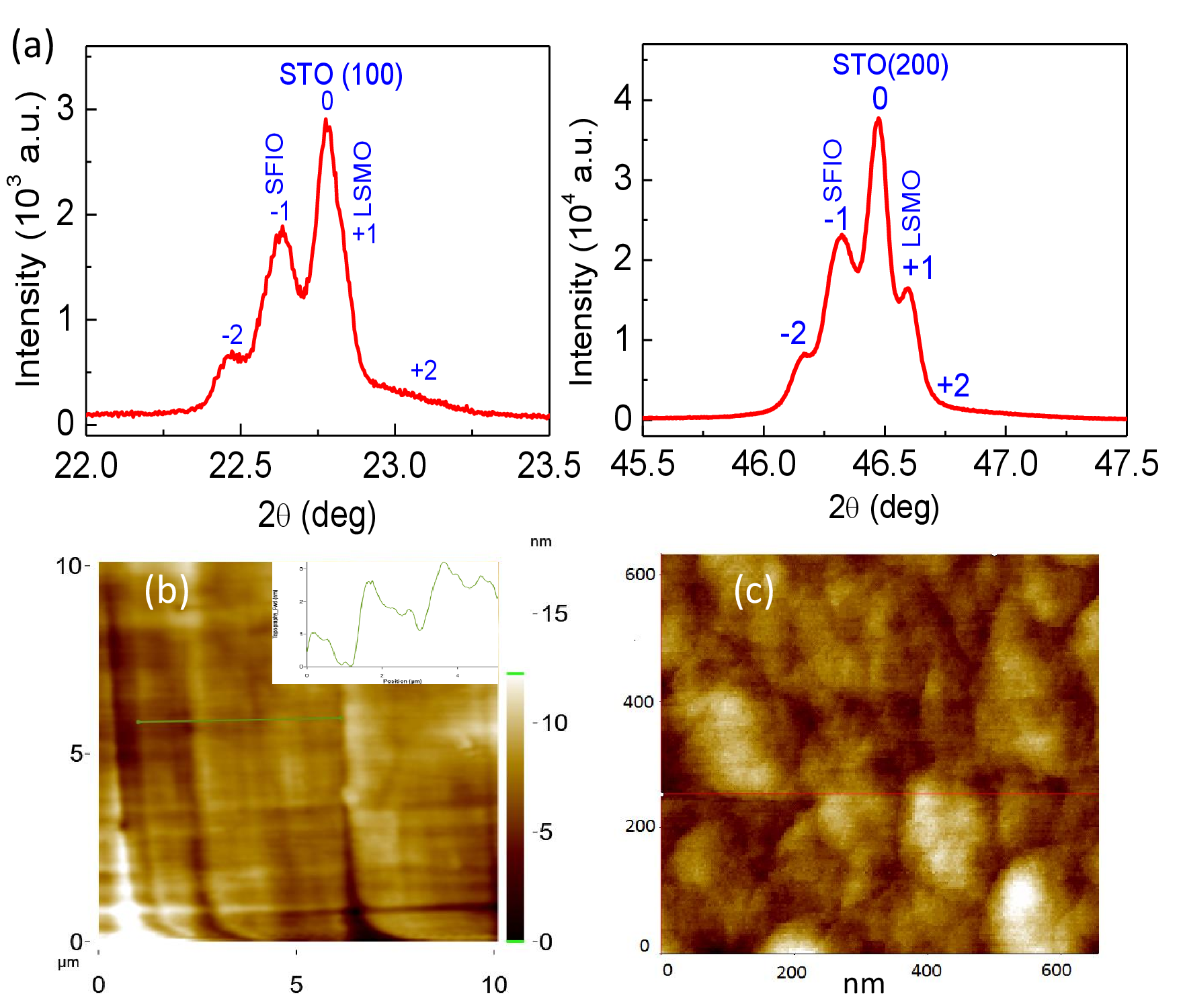}
\caption{(color online) (a) shows x-ray diffraction pattern of SL$_{10/10}$ multilayer for (100) and (200) plane at left and right panel, respectively. (b) and (c) show the atomic force microscope (AFM) image of SL$_{10/10}$ multilayer with large area (unit micro-m) and small area (unit nano-m) scanning, respectively.}
	\label{fig:Fig1}
\end{figure}

\section{Experimental details}
Multilayers of Sr$_2$FeIrO$_6$ (SFIO) and La$_{0.67}$Sr$_{0.33}$MnO$_3$ (LSMO) have been deposited on SrTiO$_3$ (100) single crystal substrate using pulsed laser deposition (KrF, 248 nm) technique with laser energy density $\sim$ 1.3 J/cm$^2$ and frequency 5 Hz. Substrate to target distance is kept around 5 cm and the deposition of both films has been done at 700${^\circ}$C to get good quality film. While deposition, oxygen pressure is maintained at $\sim$ 0.1 mbar. After deposition the chamber is filled with oxygen at pressure around 500 mbar and then normally cooled to room temperature. The Polycrystalline target materials of SFIO and LSMO have been prepared by solid state method and characterized with x-ray diffraction (XRD). First, a layer of LSMO is deposited on SrTiO$_3$ and then SFIO is deposited. This has been repeated three times to get multilayer of the form [SFIO/LSMO]$_3$. Two sets of multilayer are deposited, namely SL$_{10/10}$ and SL$_{10/5}$ where the thickness of SFIO/LSMO has been kept as 10/10 and 10/5 nm, respectively. The growth rate for film deposition has been determined using a test film which is deposited using 8000 shots of laser with above mentioned parameters. A step has been introduced while deposition of film using a clip where the step height corresponds to the thickness of film. A field-emission scanning electron microscope (FESEM) has been used to measure the thickness of deposited film while a thickness profilometer and an atomic force microscope (M/s Nanomagnetics) has been used to measure the depth of introduced step in film. Crystalline quality of the film has been checked with x-ray diffraction (XRD) measurements (PANalytical X'pert PRO) using Cu K$_\alpha$ x-ray source. Surface morphology of the film has been checked using atomic force microscope Magnetization measurements have been done using a vibrating sample magnetometer (PPMS by Quantum Design).

\begin{figure}
\centering
		\includegraphics[width=8cm]{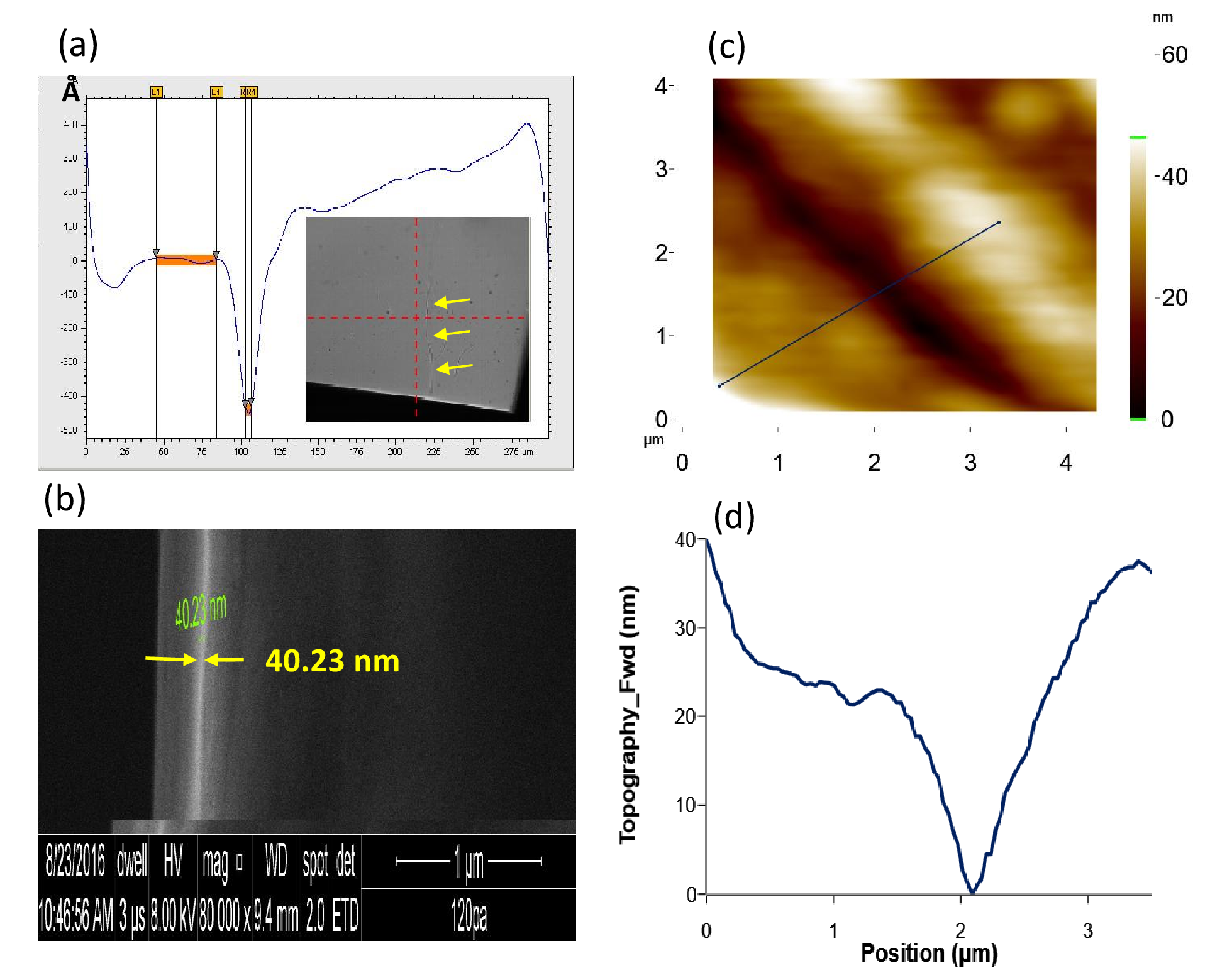}
\caption{(color online) (a) shows depth profile of single layer thin film with film thickness around 40 nm. Inset shows surface of the film while horizontal arrows indicate the step in the film indroduced during deposition of the film. (b) shows cross-sectional FESEM image same thin film showing film thickness $\sim$ 40 nm. (c) shows AFM image of the same film collected across the step.(d) shows line profile of AFM image indicating depth of the step around 40 nm.}
	\label{fig:Fig2}
\end{figure}   

\section{Result and discussion}
Fig. 1a shows XRD pattern for SL$_{10/10}$ where the data at left and right panel represents (100) and (200) planes, respectively. The films are found to be epitaxial, adopting the structure of substrate. Figure shows both superlattice peaks and (weak) thickness fringes for both peaks, where the XRD peaks for SFIO and LSMO are observed at left and right side of STO, respectively (marked as $-1$ and $+1$). Lattice parameters of SFIO and LSMO, SrTiO$_3$ are calculated from the XRD pattern. The lattice parameters for SFIO, LSMO  are 3.91 {\AA}, 3.88 {\AA} respectively, whereas 3.9 {\AA} for SrTiO$_3$. Similar, XRD pattern has been observed for SL$_{10/5}$, where peaks become less broaden (Not shown). Lattice mismatch between substrate and film has an interesting effect which induces strain and modifies the physical properties accordingly.\cite{Hwang} The lattice mismatch ($\Delta a$) has been calculated from XRD pattern using the formula \\

\begin{eqnarray}
   \Delta a (\%) = \frac{a_s -a_{th}}{a_s} \times 100
\end{eqnarray}

where $a_s$ and $a_{th}$ are the corresponding lattice parameter of substrate and thin film. The bulk lattice parameters of SFIO are $a$ = 5.5515, $b$ = 5.5785, $c$ = 7.8435 {\AA} with triclinic crystal structure and LSMO are $a$ = 5.4820, $b$ = 5.4820, $c$ =13.4490 {\AA} with rhombohedral crystal structure.\cite{Khark} Here, the lattice mismatch between SrTiO$_3$ and first mono layer i.e LSMO is around 0.51\% where the film is supposed to be grown along \textit{a-b} plane with \textit{c}-axis of the thin film contracted. However, with successive growth lattice mismatch at the interfaces of LSMO and SFIO is  0.77\%, where due to lattice mismatch \textit{c}-axis of the film has been elongated.The thickness of whole multilayer ($D$) has been calculated using the position of superlattice peaks with following formula,\cite{Kw}

\begin{eqnarray}
   D = \frac{(m-n)\lambda}{2(\sin{\theta_m} - \sin{\theta_n})}
\end{eqnarray}

where, $\lambda$ is the wavelength of x-rays used for XRD measurements, $\theta_{m}$ and $\theta_{n} $ are the position of $m$ and $n$ order peak. The $D$ for (100) and (200) planes has been calculated to be around 54.9 and 61.4 nm, respectively, where both these values of $D$ are close to expected total thickness (60 nm) of the present multilayer.

Atomic force microscope images of the multilayer SL$_{10/10}$ are shown in Fig. 1b and 1c with large and small scan area, respectively. Fig. 1b shows terrace type growth following the surface of STO substrate where inset indicates the terrace height is around 2 nm. The small area scanning in Fig. 1c shows growth pattern of present multilayer. The root mean square (rms) roughness of the film has been found to be about 1.8 nm, which is suggestive of good quality of film.

\begin{figure}
\centering
		\includegraphics[width=8cm]{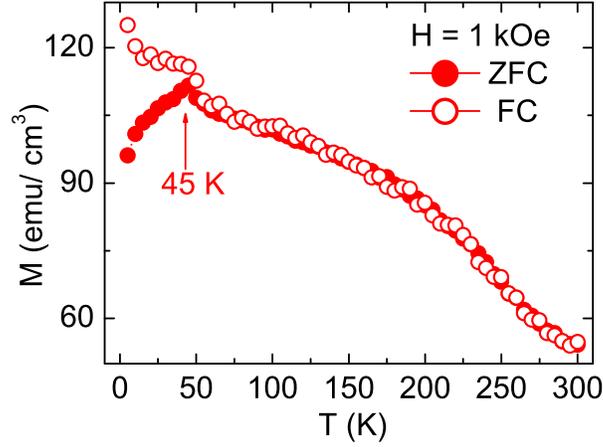}
\caption{(color online) DC magnetization data measured in 1 kOe applied field under zero field cooled (ZFC) and field cooled (FC) protocol are shown as a function of temperature for SL$_{10/10}$ multilayer. }
	\label{fig:Fig3}
\end{figure}

Fig. 2a shows the depth profile plot, collected with profilometer, on surface of test film (deposited with 8000 laser shots) across the introduced step which indicates step height is about 41 nm. A cross-sectional FESEM image of film has been shown in Fig. 2b indicating thickness of film around 40.2 nm. Fig. 2c presents an atomic force microscope image on film surface across the step. The line profile of microscope image has been shown in Fig. 2d which implies step height $\sim$ 40 nm. The Fig. 2, as a whole, suggests the average thickness of film deposited with 8000 laser shots is around 40 nm which has been used for growth rate calibration.

Fig. 3 shows temperature dependent magnetization $M(T)$ data measured in 1000 Oe following zero field cooled (ZFC) and field cooled (FC) protocol for SL$_{10/10}$. With cooling, both $M_{ZFC}$ and $M_{FC}$ increases following the magnetic nature of LSMO which develops FM ordering at temperature above the room temperature. However, at low temperature around 45 K the $M_{ZFC}$ exhibits a kink and below this temperature a wide bifurcation is observed between the ZFC and FC magnetization data. It can be noted that bulk Sr$_2$FeIrO$_6$ exhibits an AFM transition and onset of bifurcation between ZFC and FC $M(T)$ data around 45 K ($T_N$).\cite {Khark} Similar AFM type magnetic transition at same temperature has been observed in Sr$_2$FeIrO$_6$ thin film (Fig. 3), although the nature of bifurcation in present multilayer is quite different than bulk Sr$_2$FeIrO$_6$. This kink in $M_{ZFC}$ and bifurcation is likely due to an interface effect. In particular, the decrease of moment in ZFC data below 45 K suggests an AFM type spin coupling at the interface. On the other hand, increase of moment below $T_N$ in FC data is probably due to an increase of FM type interaction at interface which is favored while cooling the sample in magnetic field.

\begin{figure}
\centering
		\includegraphics[width=8cm]{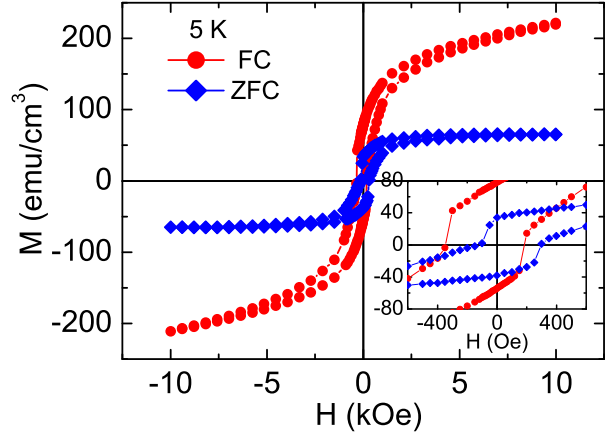}
\caption{(color online) Field dependent magnetization data collected at 5 K following zero field cooled (ZFC) and field cooled (FC) protocol  are shown for SL$_{10/10}$ multilayer. Inset shows expanded view of the hysteresis loop near origin.}
	\label{fig:Fig4}
\end{figure}

The field dependent magnetization data $M(H)$ collected following ZFC and FC protocol at 5 K with magnetic field range $\pm$ 10 kOe ($H_{max}$) are shown in Fig. 4 for SL$_{10/10}$ multilayer. For ZFC data, the sample is cooled from room temperature to the target temperature in zero magnetic field and then $M(H)$ data have been collected. We have varied maximum applied field $H_{max}$ in ZFC measurements at 5 K. For FC $M(H)$ data, the system has been cooled in presence of magnetic field ($H_{cool}$) to the target temperature and then field is swept from +$H_{cool}$ to -$H_{cool}$ and then to +$H_{cool}$. For both ZFC and FC $M(H)$ data, substrate (diamagnetic) contribution has been corrected using a slope in $M(H)$ data taken at high field regime of ZFC $M(H)$. The moment value in all $M(H)$ data represents contribution from both SFIO and LSMO layers. As evident in Fig. 4, magnetic moment in ZFC $M(H)$ initially increases steeply till $ H \sim$ 1 kOe and then the increase of $M(H)$ is rather slow. This initial fast increase of $M(H)$ is probably due to LSMO which is soft FM and shows saturation above $\sim$ 1 kOe.\cite {Feng} The SFIO is a hard AFM, therefore, ZFC $M(H)$ shows a slow increase or nearly saturation around 1 kOe field.\cite{Khark} FC $M(H)$ data, on the other hand, continuously increase and it has much higher value compared to ZFC $M(H)$. This further indicates an AFM type spin interaction between LSMO and SFIO at interface. During FC process, the magnetic field couples with the interfacial spins while cooling through $T_N$ of SFIO and induces FM like exchange which basically softens the spin alignment. Therefore, $M(H)$ shows a continuous increase in higher field regime.

\begin{figure}
\centering
		\includegraphics[width=8cm]{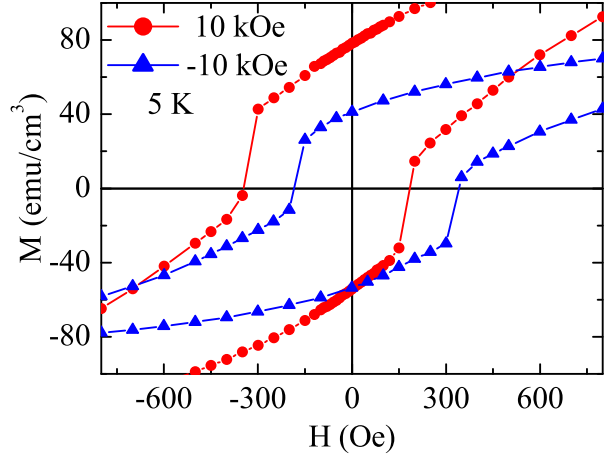}
\caption{(color online) Field dependent magnetization data collected at 5 K following field cooled (FC) protocol with cooling field +10 kOe and - 10 kOe respectively are shown for SL$_{10/10}$ multilayer. }
	\label{fig:Fig5}
\end{figure}

Interestingly, both ZFC and FC $M(H)$ data show an asymmetry i.e., $M(H)$ loop is shifted (inset of Fig. 4). We observe $H_c^L$ and $H_c^R$ = - 346.1 and 183.8 Oe, and $M_r^U$ and $M_r^L$ = 78.3  and -54.2 emu/cc, respectively. However, the direction of shifting is opposite in ZFC $M(H)$ data which moves to positive field and negative moment axis. We find $H_c^L$ = - 127.4 and $H_c^R$ = 296.4 Oe, and $M_r^U$ = 34.0 and $M_r^L$ = - 38.3 emu/cc. The $H_c^L$ and $H_c^R$ are the left and right coercive field, respectively where the moment becomes zero. Similarly, $M_r^U$ and $M_r^L$ are the upper and lower moment value, respectively at $H$ = 0. The nature of shift as well as closing of loops implies an EB effect in present multilayer. We have calculated the exchange bias field ($H_{EB}$) and effective coercive field $H_c$ using $H_{EB}$ = ($|H_c^L$ + $H_c^R|$)/2 and $H_c$ = ($|H_c^L$ - $H_c^R|$)/2, respectively. For FC $M(H)$, we find $H_{EB}$ = 81.2 and $H_c$ = 265 Oe, respectively. Similarly, ZFC $M(H)$ data give $H_{EB}$ = 84.5 and $H_c$ = 211.9 Oe, respectively. It is though surprising that EB effect has been observed in both ZFC and FC $M(H)$ data in same system which is very uncommon. It can be noted that exchange bias has not been observed in single layer thin film of LSMO and SFIO deposited on SrTiO$_3$ substrate (Not shown). Conventionally, magnetic loop shifting occurs due to an unidirectional FM anisotropy when FM/AFM interface is cooled in magnetic field. The $M(H)$ loop generally shifts in opposite direction of cooling field i.e., shifts toward negative field when cooled in positive field which is called as negative EB effect.\cite{Goke,Nogues,Nog} It is further believed that FM type spin exchange coupling at FM/AFM interface renders an unusual EB effect. However, there are only few studies which have reported $M(H)$ loop shifting, normally toward positive field axis, even cooling in zero magnetic field which is known as positive or spontaneous EB effect. An AFM type spin interaction is argued for this positive EB effect which has been observed only in few systems.\cite{Wang,Saha,Murthy} Nonetheless, this ZFC EB effect is quite intriguing as the prerequisite unidirectional anisotropy, which is otherwise induced during FC process, can be realized isothermally in ZFC process.

\begin{figure}
\centering
		\includegraphics[width=8cm]{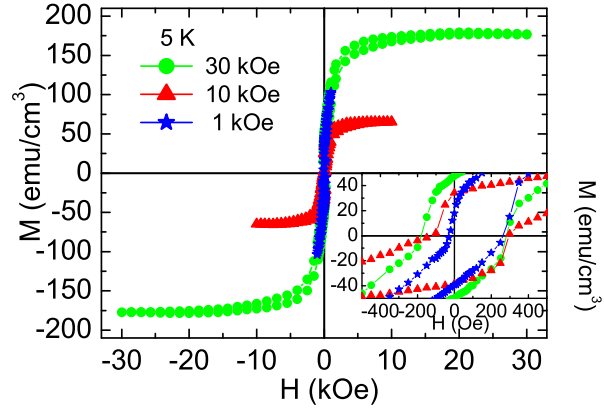}
\caption{(color online) Field dependent magnetization data collected at 5 K  following zero field cooled (ZFC) protocol are shown with maximum field 1 kOe, 10 kOe and 30 kOe respectively, for SL$_{10/10}$ multilayer. Inset shows expanded view of the hysteresis loop near origin.}
	\label{fig:Fig6}
\end{figure}

To confirm the exchange bias effect in present multilayer, we have collected FM $M(H)$ data after cooling in both positive and negative applied field at 5 K. Fig. 5 shows $M(H)$ hysteresis loop close to the origin for cooling field of +10 and -10 kOe field. As evident in figure that for positive field cooling the $M(H)$ shifts to negative field axis and for negative field cooling the $M(H)$ shows opposite shifting. However, the shifting of $M(H)$ is almost equal. We calculate the $|H_{EB}|$ around 81 Oe in both cases. This confirms the exchange bias phenomena in present multilayer. 
 
\begin{figure}
\centering
		\includegraphics[width=8cm]{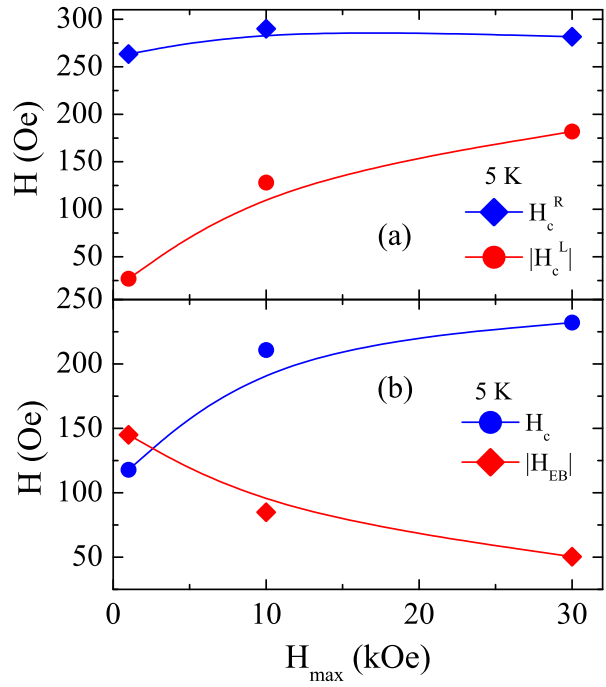}
\caption{(color online) (a) Right ($H_c^R$) and left ($H_c^L$) coercive fields are shown as a function of maximum applied field ($H_{max}$) that are obtained in ZFC $M(H)$ measurements for SL$_{10/10}$ multilayer at 5 K. (b) shows calculated coercive ($H_c$) and exchange bias ($|H_{EB}|$) field as a function of maximum field at 5 K (see text). }
	\label{fig:Fig7}
\end{figure}

In order to understand the ZFC EB effect in further detail, we have measured $M(H)$ loops at 5 K with different maximum applied field $H_{max}$ for SL$_{10/10}$. Fig. 6 shows $M(H)$ plots with $H_{max}$ = 1, 10, and 30 kOe at 5 K. As evident in figure, while $M(H)$ data for $H_{max}$ = 10 and 30 kOe show almost saturation, the data for $H_{max}$ = 1 kOe show no sign of saturation. Inset of Fig. 6 shows expanded view of same plot near the origin. It is surprising that the moment in ZFC $M(H)$ increases substantially with applied field, $H_{max}$. Out of Mn, Fe and Ir transition metals, only Mn (mixture of Mn$^{+3}$ and Mn$^{+4}$) and Fe$^{+3}$ contribute to moment as Ir$^{+5}$ is supposed to be nonmagnetic.\cite{Khark} This implies the saturation moment for LSMO and SFIO would be 3.7 and 5 $\mu_B$/f.u., respectively. We, however, experimentally find moment around 0.807 and 2.193 $\mu_B$/f.u. in ZFC $M(H)$ for 10 and 30 kOe field (Fig. 6), respectively considering only LSMO layers. This indicates even FM LSMO layer is not fully saturated which is probably due to surface/interface disorder and finite-size effect in films. It has been previously shown that both the transition temperature $T_c$ as well as moment is significantly modified with layer thickness in LSMO film.\cite{Kour, Huij, Chop} The Figure 6 suggests that initially with increasing field the moment in LSMO layers and at interfaces increases which gives different values of moment at different fields. However, the moment in AFM type SFIO layer does not increase much, and as a resultant it gives almost saturated value in higher field regimes. 

All the $M(H)$ plots show shifting toward positive field axis i.e., ZFC EB effect, however, nature of shifting changes with $H_{max}$. For $H_{max}$ = 1 kOe, we find $H_c^L$ and $H_c^R$ values are around 27 and 263 Oe, and $M_r^U$ and $M_r^D$ values are about 17.8 and 39.3 emu/cc, respectively. While though $M(H)$ does not saturate with $H_{max}$ = 1 kOe but the $H_c^L$ closely matches with that for LSMO films.\cite {Ett, Feng, Kour, Huij, Chop} The high value of $H_c^R$ probably arises due to locking of spin at the interface of LSMO/SFIO which causes positive EB effect. With increasing $H_{max}$, Fig. 7 shows while $H_c^R$ does not change appreciably but $H_c^L$ increases significantly and tends to approach $H_c^R$. In Fig. 7a, we have shown the variation of both $H_c^R$ and $H_c^L$ with $H_{max}$ at 5 K for SL$_{10/10}$ multilayer. The calculated $H_{EB}$ and $H_c$ are shown in Fig. 7b which shows $H_{EB}$ decreases and $H_c$ increases with $H_{max}$.

\begin{figure}
\centering
		\includegraphics[width=8cm]{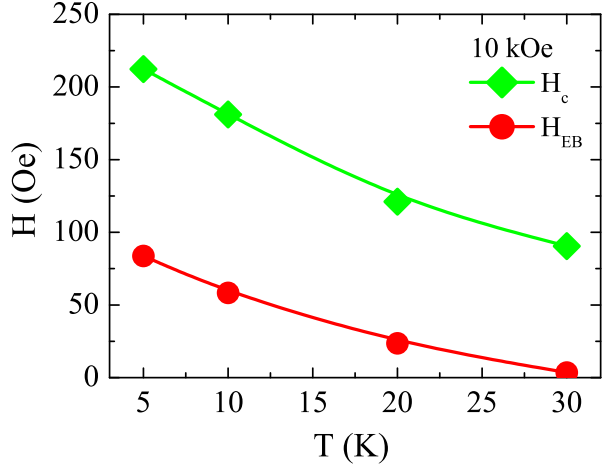}
\caption{(color online) Variation of coercive and exchange bias field is shown with temperature in zero field cooled (ZFC) field dependent magnetization for SL$_{10/10}$ multilayer.}
	\label{fig:Fig8}
\end{figure}
    
We have further checked the effect of temperature on EB effect. We have measured ZFC $M(H)$ loop with $H_{max}$ = 10 kOe at different temperatures for SL$_{10/10}$ multilayer. The calculated $H_{EB}$ and $H_c$ are presented in Fig. 8 showing both the values decrease monotonically with temperature. The $H_{EB}$ almost vanishes ($\sim$ 4 Oe) once temperature is raised to 30 K which is close to $T_N$ ($\sim$ 45 K) of AFM component SFIO. With increasing temperature, the AFM anisotropy in SFIO weakens which consequently reduces the EB effect. 

In Fig. 9, we propose a schematic model for spin alignment with magnetic field of SL$_{10/10}$ multilayer during ZFC $M(H)$ process. It is clear in Fig. 6 that the externally applied $H_{max}$ has significant influence on the the spin interaction in present multilayer system. For example, asymmetric behavior in $M(H)$ is reduced where the $H_c^L$ increases which gives a decreasing exchange bias effect with increasing $H_{max}$ (see Fig. 7). Given that two magnetic component in present multilayer are FM (LSMO) and AFM (SFIO) and it shows different EB in ZFC, therefore an AFM type interface exchange interaction expected. In Fig. 9, we have focused on a representative tri-layer of LSMO/SFIO/LSMO (FM/AFM/FM) showing two interfaces. While there are many domains of different spin orientation in each layer, for simplicity, we have considered only two domains (Region I and II) in each layer with different crystal easy axis in parallel and anti-parallel to applied field. It can be further noted that there are very limited reports of neutron diffraction measurements at low temperature magnetic state (5 K) for SFIO to understand the AFM spin structure or the nature of magnetic exchange interaction.\cite{battle,qasim} The study shows a nearest neighbor type exchange interaction among magnetic Fe$^{+3}$ ions but the nature of AFM type has not been discussed. In Fig. 9, we have assumed a simple AFM spin structure where an alternative layer has opposite orientation which is close to A-type AFM. This has been shown to schematically demonstrate the interface exchange interaction. Nonetheless, experimental efforts are required to understand the AFM spin structure in SFIO.

Fig. 9a shows the tri-layer assembly with an AFM type interface at 5 K under ZFC condition. When the applied magnetic field during first application of magnetic field in $M(H)$ measurement is low and below the threshold value $H_{Th}$, the LSMO spins (Region II) will align to the direction of magnetic field due to its low anisotropy (Fig. 9b). The antiferromagnetically ordered spins in hard SFIO will, however, not be influenced at low magnetic field. This LSMO spin alignment will reconstruct the AFM-type interface in Region II (Fig. 9b). With further increase in magnetic field above $H_{Th}$, the SFIO spins in the vicinity of interface will align to the direction of magnetic field because at interface generally has low anisotropy compared to bulk of material. This spin alignment conversion will advance the interface toward inside of AFM SFIO layer which would effectively increase the thickness of FM layer. More the applied field, less would be the separation between FM layer as shown at Fig. 9c. When the applied magnetic field $H_{max}$ is sufficiently high, an effective thickness of AFM layer will reduce substantially which will facilitate an indirect or tunneling exchange interaction between the adjacent FM layers due to their inter-layer separation.\cite{Wang} This tunneling exchange is induced by applied magnetic field and would help to retain the moment even field decreases to zero. This tunneling exchange is similar to exchange interaction as observed in cases of cluster glass or super-ferromagnetic interaction.\cite{Bedanta} While the AFM spin ordering in SFIO will be the least influenced by field sweeping but there tunneling exchange among the FM layers will obstruct the rotation of FM spin in LSMO with field, which will result in an increased $M_r^U$ and $H_c^L$. 

The effect of tunneling FM exchange is evident in Fig. 6 as with increasing $H_{max}$, both the remnant moment $M_r^U$ as well as the coercive field $H_c^L$ increases. The decreasing exchange bias effect ($H_{EB}$) with increasing $H_{max}$ (Fig. 7) can be explained with an indirect FM coupling among  the FM layers in present multilayer system. In present model, we have considered only FM/AFM domains with anisotropy axis parallel and/or anti-parallel to the applied magnetic field for simplicity. However, domains with easy axis making finite angle with magnetic field will also result in exchange bias effect.\cite{Wang} Here, it can be mentioned that tuning of ZFC exchange bias effect through applied magnetic field has been previously observed in Ni-Mn-In bulk alloys where the superparamagnetic domains embedded in AFM host are shown to grow with the field which engage in tunneling exchange interaction that modifies both the coercive fields and consequently exchange bias effect.\cite{Wang} The remnant magnetization $M_r^U$ as well as $H_c^L$ increase with $H_{max}$ which implies a new complex magnetic state has been established in this multilayer which has altered the EB effect. The significant aspect of present SL$_{10/10}$ multilayer is that it exhibits EB effect with reasonable $H_{EB}$ in a simplified ZFC condition. Moreover, it requires low magnetic field which is in contrast with other reported studies where higher magnetic fields have been applied.\cite{Wang,Saha} Nonetheless, tuning of exchange bias effect in present heterostructure system with magnetic field is quite noteworthy.

\begin{figure}
\centering
		\includegraphics[width=8.5cm]{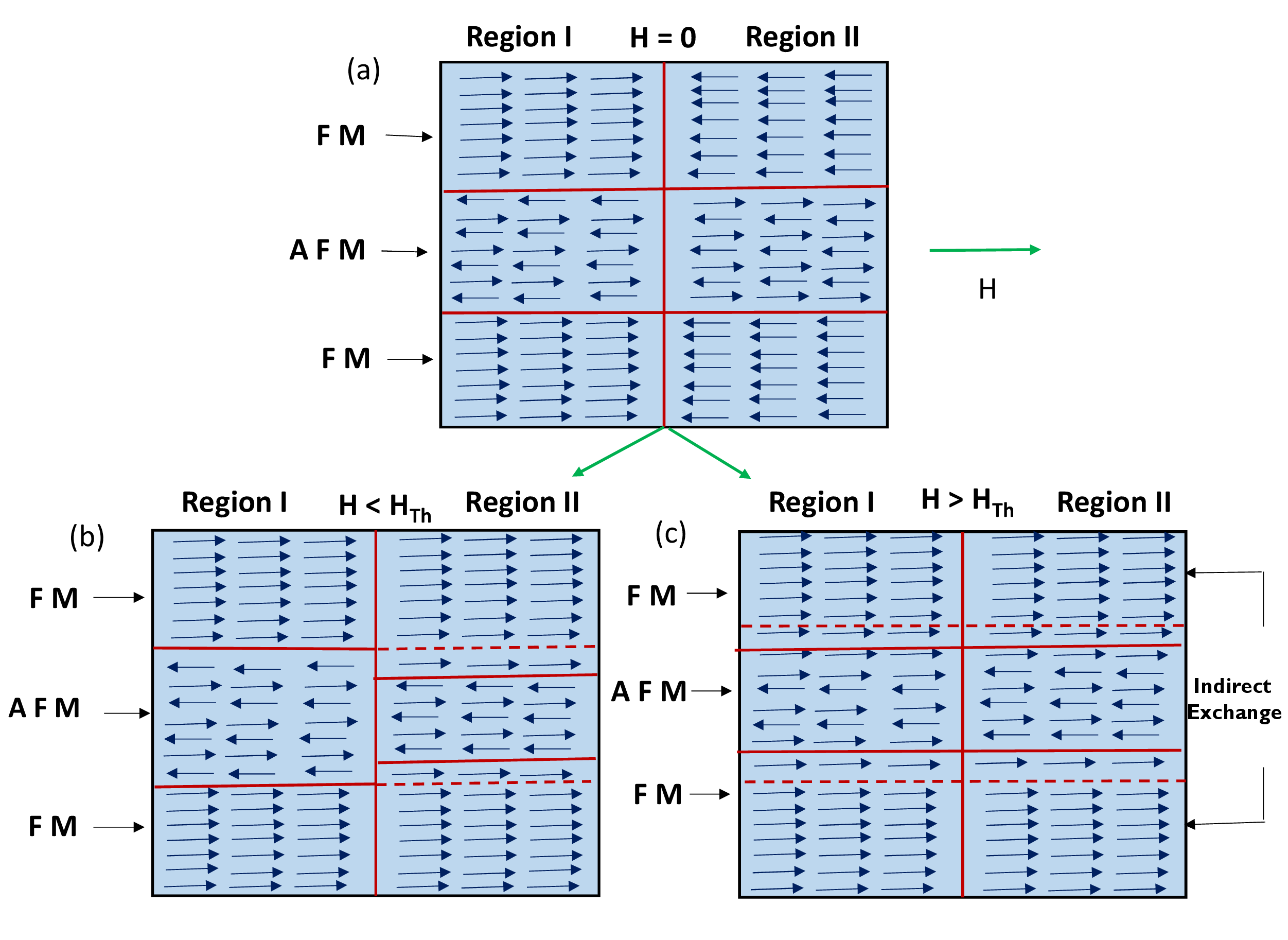}
\caption{(color online) Schematic diagram representing the spin interaction during ZFC M(H) measurement is shown at 5 K for SL$_{10/10}$ multilayer. (a) shows spin arrangement at $H = 0$ before application of magnetic field while (b) and (c) show the same at $H < H_{Th}$ and $H > H_{Th}$, respectively (see text). Vertical lines separate two region with opposite anisotropy axis, the bold horizontal lines represent FM/AFM interface while broken horizontal lines represent the position of original interface in this multilayer.} 
	\label{fig:Fig9}
\end{figure}
 
To further understand the role of individual FM and AFM layer on EB effect, we have prepared another [SFIO/LSMO]$_3$ multilayer SL$_{10/5}$ keeping layer thickness of SFIO and LSMO around 10 and 5 nm, respectively. Fig. 10 shows both ZFC and FC $M(H)$ plots of SL$_{10/5}$ at 5 K. Unlike the SL$_{10/10}$ multilayer, ZFC $M(H)$ data of SL$_{10/5}$ surprisingly do not show any sign of saturation at high magnetic field, rather it increases continuously till measuring field of 10 kOe and shows higher value than FC $M(H)$ (Figs. 4 and 10). The expanded view of $M(H)$ data close to the origin has been shown in inset of Fig. 10. With stark contrast, ZFC $M(H)$ of SL$_{10/5}$ does not exhibit any appreciable asymmetry in terms of magnetic field of magnetization i.e., it does not show EB effect. However, reasonable negative shifting of $M(H)$ data is observed for FC $M(H)$. We find $H_c^L$ = -1048.5 and $H_c^R$ = 617.0, and $M_r^U$ = 61.1 and $M_r^D$ = -44.2 emu/cc which gives $H_{EB}$ = 215.7 and $H_c$ = 832.7 Oe. These (FC) values of both $H_{EB}$ and $H_c$ are much higher compared to previously discussed SL$_{10/10}$ multilayer at same temperature and with same $H_{max}$. The effect of FM layer thickness on EB effect has been discussed theoretically which predicts $H_{EB}$ $\propto$ 1/t$_{FM}$, where $t_{FM}$ is FM layer thickness.\cite{Nogues,Kiwi} With decreasing $t_{FM}$, the FM spins are more firmly locked with AFM layer through interface exchange coupling therefore, FC spin rotation shows higher $H_c^L$ during field sweeping in M(H).  In that sense, the increase of both $H_{EB}$ and $H_c$ in SL$_{10/5}$ multilayer is quite explained through decrease of FM layer thickness. Disappearance of EB effect in ZFC $M(H)$ with decreasing FM layer thickness in SL$_{10/5}$ is quite intriguing. The ZFC $M(H)$ for SL$_{10/5}$ shows $H_c^L$ $\sim$ $H_c^R$ $\sim$ 430 Oe which are higher than the respective values for SL$_{10/10}$. Though there would be an expansion of FM layer thickness during application of initial magnetic field in ZFC $M(H)$ but the $t_{FM}$ will still be thin enough where the FM spins are firmly locked with adjacent AFM spins through interface. This would render a symmetric ZFC $M(H)$ without EB effect. Nonetheless, presence of both ZFC and FC EB effect in same system with reasonable bias field $H_{EB}$ in low magnetic field is quite noteworthy and requires further investigation using both theoretical and microscopic experimental tools. 

\begin{figure}
\centering
		\includegraphics[width=8cm]{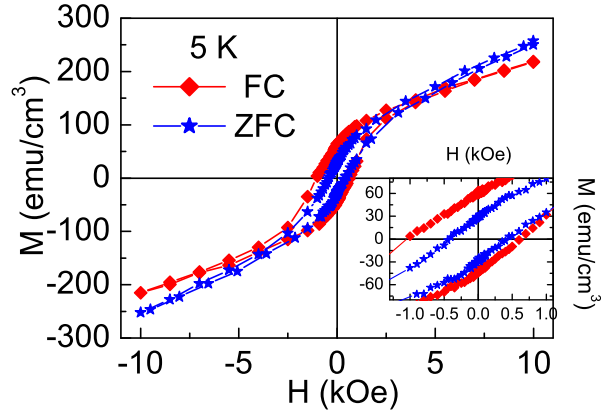}
\caption{(color online) Zero field cooled (ZFC) and field cooled (FC) magnetization data are shown as a function of magnetic field collected at 5 K for SL$_{10/5}$ multilayer. Inset shows expanded view of the hysteresis loop near origin.}
	\label{fig:Fig10}
\end{figure}

\section{Conclusions}
Epitaxial multilayers of 3$d$-5$d$ based La$_{0.67}$Sr$_{0.33}$MnO$_3$ and double perovskite Sr$_2$FeIrO$_6$ have been deposited on single crystal SrTiO$_3$ (100) using pulsed laser deposition technique. The magnetic states of bulk La$_{0.67}$Sr$_{0.33}$MnO$_3$ and Sr$_2$FeIrO$_6$ are respectively ferromagnetic with $T_c$ above room temperature and AFM with ordering temperature $T_N$ $\sim$ 45 K. An onset of antiferromagnetic type magnetic exchange coupling at interface is evident in temperature dependent magnetization data below $T_N$ of Sr$_2$FeIrO$_6$. Interestingly, multilayer with $\sim$ 10 nm of individual layer thickness exhibits both FC and ZFC exchange bias effect. While the EB effect after FC is conventional but the positive EB effect after ZFC is quite unusual and rarely observed. In ZFC $M(H)$, it is believed that FM layer thickness increases with field and a tunneling exchange coupling is established among the FM layers which basically weakens the ZFC EB effect with increasing maximum applied field. Similarly, ZFC EB effect decreases with increasing temperature which is due to weakening of AFM anisotropy. The 3$d$-5$d$ interface is quite interesting due to presence of different competing energy scale. Therefore, more similar studies involving different transition metals are necessary to comprehend this complex behavior.       

\ack{We acknowledge SERB, DST for supporting Excimer pulse laser and UPE-II, UGC for deposition chamber. We are also thankful to DST-FIST for supporting atomic force microscope, DST-PURSE for funding and IIT Delhi for magnetization measurements. KCK and RC acknowledge UGC, India for financial support.}

\end{document}